\documentstyle[preprint,aps]{revtex}

\def\twen{\frac {1}{27}}
\def\eiphi{e^{i\phi}}
\def\diff{(1-\eiphi)}
\def\one{(-\twen+\frac {\diff}{35})}
\def\two{-\frac {\diff}{35}}
\def\three{(-\twen+\frac {\diff}{45})}
\def\four{-\frac {\diff}{45}}
\def\fourp{\frac {5}{9}\diff}
\def\term1{{\frac {1} {15}} ({\frac {12}{27}}-\eiphi)}
\def\kx{\frac{5}{3}\diff}
\def\k10{{-\frac {1}{15}\diff}}
\def\five {-(\frac{2}{135}+\frac{\eiphi}{45}) }
\def\six {-\frac{\diff}{150}}
\def\sixp {\frac{2}{9}\diff F}
\def\seven {-(\frac{1}{15}+\frac{1}{10}\eiphi)}
\def\eight {\frac{8}{10}(\twen + \frac {\eiphi}{8}) }
\def\nine {-(\frac {20}{27}+\frac {5\eiphi}{2})F}
\def\ten {(\frac {-1}{45} +\frac{\eiphi}{20})}
\def\eleven {-(\frac{2}{135}+\frac{\eiphi}{20})}
\def\twelve {(\frac {10}{27}+\frac{5\eiphi}{4})F}
\def\thirteen {-\frac {\diff}{35}}
\def\fourteen {-(\frac{4}{525}+\frac {9\eiphi}{350}) }
\def\fifteen {(\frac {8}{27}+\eiphi)F}
\def\sixteen {-\frac {1}{35}\diff}
\def\seventeen {-(\frac {4}{1575}+\frac {3\eiphi}{350})}
\def\eighteen {(\frac {16}{189}+\frac {2\eiphi}{7})F}
\def\nineteen {-(\frac {1}{15}+\frac{\eiphi}{10})}
\def\twenty {(\frac {2}{225}+\frac {3\eiphi}{100})}
\def\twentyone {-(\frac {8}{27}+\eiphi)F}
\def\twentytwo {(\frac {8}{27}+\eiphi)F}
\def\tthree {(\frac {-2}{135}+\frac{\eiphi}{30})}
\def\tfour {-(\frac {1}{45}+\frac{\eiphi}{30})}
\def\tfive {\frac {-\diff}{45}}
\def\tsix {-(\frac {1}{75}+\frac{\eiphi}{50})}
\def\tseven {(\frac {1}{27}-\frac{4\eiphi}{3})F}
\def\teight {(\frac {14}{27}+\frac{7\eiphi}{9})F}
\def\tnine {\frac {-\diff}{45}}
\def\tten {-(\frac {1}{225}+\frac{\eiphi}{150})}
\def\televen {(\frac {19}{27}-\frac{\eiphi}{3})F}
\def\ttwelve {-(\frac {4}{27}+\frac{2\eiphi}{9})F}
\def\sone {-\frac {\diff}{15}}
\def\stwo {(\frac {2}{75}-\frac{3\eiphi}{50})}
\def\sthree {(\frac {73}{27}-4\eiphi)F}
\def\sfour {(\frac {-28}{27}+\frac {7\eiphi}{3})F}
\def\sfive {\frac {-\diff}{15}}
\def\ssix {(\frac {2}{225}-\frac{\eiphi}{50})}
\def\sseven {(\frac {37}{27}- \eiphi)F}
\def\ssevenp {(\frac {8}{27}- \frac {2\eiphi}{3})F}
\def\vone {-(\frac{1}{45}+\frac {\eiphi}{30})}
\def\vtwo {-(\frac{4-9\eiphi}{900})}
\def\vthree {(\frac {19}{27}+\frac {\eiphi}{2})F}
\def\vfour {(\frac {-4}{27}+\frac {\eiphi}{3})F}
\def\vfive {-(\frac{1}{75}+\frac {\eiphi}{50})}
\def\vsix {\frac{-\diff}{150}}
\def\vseven {(\frac{8}{27}+\eiphi)F}
\def\veight {-(\frac{1}{225}+\frac {\eiphi}{150})}
\def\vnine {-\frac {\diff}{150}}
\def\vten {\frac {10}{27}F}

\begin{document}

\begin{titlepage}
\begin{flushright}
NSF-ITP-96-04\\
NUB-TH-3131\\
\end{flushright}

\begin{center}
\Large{\bf Textured Minimal and Extended Supergravity Unification and 
 Implications for Proton Stability }\\
Pran Nath \\
\small{\it Institute for Theoretical Physics \\
University of California \\
Santa Barbara, CA 93106-4030 \\
\& \\
Department of Physics, Northeastern University \\
Boston, MA 02115}\footnote{Permanent address} \\
\date{\today}
\end{center}
\begin{abstract}
We  construct a class of textured supergravity  unified SU(5) models using 
Planck scale corrections. We show that the texture constraints in the Higgs 
doublet sector are insufficient in general to fully determine the textures 
in the Higgs triplet sector. A classification of textured minimal parameter 
models is given and their Higgs triplet textures computed under the constraint
that they possess the Georgi-Jarlskog textures in the Higgs doublet sector.
It is argued that additional dynamical assumptions are needed to remove 
the ambiguity.The recently proposed extension of supergravity unification 
to include a minimal exotic sector is free of this ambiguity and leads to 
unique textures in the Higgs triplet sector. Implications for proton stability 
are discussed.
 
\end{abstract}
\end{titlepage}
The concept of quark-lepton textures at the GUT scale\cite{georgi,harvey}
 has played a key role recently in the understanding of the hierarchy of 
mass scales at the electro-weak scale\cite{anderson}. Without the textures 
GUT models make poor predictions for the quark lepton mass ratios. Thus,for 
example, while the supersymmetric SU(5) model makes acceptable predictions for 
$m_b/m_{\tau}$ , the predictions of the model for the light quark-lepton 
mass ratios, i.e., $m_s/m_{\mu}$ and $m_d/m_e$ are in poor agreement with 
experiment.Supergravity grand unification\cite{can,an} currently  provides a 
successful framework for the breaking of supersymmetry.Recently, the framework 
of supergravity unification was  extended to include textures\cite{hierarchy}.
The extension
was based on the inclusion of a new sector which contains exotic matter,
which couples to matter in the visible sector and in the hidden sector.
After spontaneous breaking of supersymmetry, exotic matter becomes 
superheavy and its elimination leads to a well defined set of higher 
dimensional operators scaled by $\Sigma/M_P$, where $\Sigma$ is the 
24-plet of SU(5). Textures are created when SU(5) breaks to SU(3)x
SU(2)xU(1) at the GUT scale. It is then shown that if one fixes the 
textures in the Higgs doublet sector, then the textures in the Higgs
triplet sector are uniquely determined. 

	In this Letter we consider  a more general approach.Here instead
of generating higher dimensional operators via  the exotic sector, we 
add in a phenomenological fashion a set of higher dimensional operators.
That higher dimensional operators can generate hierarchies in quark-lepton 
mass matrices has been known for some time\cite{froggatt} and further 
one expects such operators to arise quite naturally in string compactified 
models\cite{cchl,dienes}. We shall show that in this 
case the constraints that fix the textures in the Higgs doublet
sector leave a considerable degree of arbitrariness in the textures 
in the Higgs triplet sector. We then classify the minimal parameter solutions
and find that there are at least  4x5x17 textured models of this type
( which we label by $A_iB_jC_k$ (i=1,..4;~j=1,..,5;~k=1,..,17) which
 posess the same Georgi-Jarlskog(GJ) textures in the Higgs doublet sector but 
 have distinct textures in the Higgs triplet sector. We compute the textures
in the Higgs triplet sector for these $4\times 5\times 17$ minimal 
parameter models.They are given by eqs(9), (12) and (13) and tables 1,2 and 
4. These results have 
important implications for p-decay lifetimes\cite{wein,acn}.

We give now the details of the analysis. 
As discussed above textures in the quark-lepton sectors can arise 
via higher dimensional operators.For the minimal SU(5) theory these
 higher dimensional operators are scaled by $\Sigma/M_P$.
The hierarchy of mass scales arises when the 24-plet of $\Sigma$ field 
develops a VeV generating the ratio $M/M_P$ ,where M is 
the GUT scale and $M_P$ is the Planck/string scale.
 As is commonly done we shall assume  that the (33) element of the up quark
 texture arises from a dimension four operator in the Lagrangian 
(or dimension 3 in the superpotential) while the 
remaining parts of the up quark texture and all parts of the down quark and 
lepton textures arise from interactions with dimensionalities higher 
than four. To generate the full hierarchical structure one has to 
include up to dimension six operators in the up quark sector and up to 
dimension seven operators in the down quark and lepton sector.
As discussed above we shall take a phenomenological approach and
write down the  general set of interactions at each level of 
dimensionality with only the constraint of R-parity invariance.
In general, the interaction structure will have the form 

\begin{equation}
W=W_3+W_4+W_5+W_6+..
\end{equation}
 We assume that the particle spectrum is that 
of the minimal supersymmetric SU(5) model, and consists of quarks and leptons 
in three  generations of $\bar 5 (M_x)+ 10(M^{xy})$ plets of SU(5) , 
Higgs in $\bar 5(H_{1x}) +5(H_2^x)$, and a  
field $\Sigma_y^x$ that breaks the SU(5) GUT symmetry in the 24-plet of SU(5). 
In the computation of textures in the up quark sector it is found sufficient 
to include only the first three terms of the expansion on the right hand side
of eq(1), i.e., the terms $W_3,W_4,W_5$, to generate the desired 
hierarchies and $W_6$ and higher terms make small contributions and can be 
neglected.In the down quark and lepton sector we assume that $W_3$ makes 
no contribution and  $W_4,W_5,W_6$ are then found sufficient to generate the 
desired hierarchies and $W_7$ and higher terms can be neglected.
Under the above conditions the desired interactions are given by 

\begin{equation}
W_3=-\frac{1}{8}\epsilon_{uvwxy}H_2^uM_i^{vw}h_{ij}M_j^{xy} 
+ H_{1x}M_{yi}k_{ij}M_j^{xy}
\end{equation}

\begin{eqnarray}
W_4
&=&-\frac{1}{8M_P}\epsilon_{uvwxy}\Sigma_{q}^{u} H_2^q M_i^{vw}h_{1ij}M_j^{xy}
-\frac{1}{8M_P}\epsilon_{uvwxy}\Sigma_q^{u} M_i^{qv}H_2^w h_{2ij}M_j^{xy}
\nonumber \\
& &\mbox{}+\frac {1}{M_P} H_{1x}\Sigma_y^x M_{zi}k_{1ij}M_j^{yz}
+ \frac {1}{M_P} H_{1x}M_{yi} \Sigma_z^y k_{2ij}M_j^{xz}
\end{eqnarray}
\begin{eqnarray}
W_5=-\frac{1}{8M_P^2}\epsilon_{uvwxy}\Sigma_{q}^{2u} H_2^q M_i^{vw}h_{3ij}
M_j^{xy}
-\frac{1}{8M_P^2}\epsilon_{uvwxy}\Sigma_q^{2u} M_i^{qv}H_2^w h_{4ij}M_j^{xy}
\nonumber\\
\mbox{}-\frac{1}{8M_P^2}\epsilon_{uvwxy}\Sigma_{q}^{u} H_2^q M_i^{vw}h_{5ij}
\Sigma_z^x M_j^{zy}
\mbox{}-\frac{1}{8M_P^2}\epsilon_{uvwxy}H_2^u \Sigma_{v}^{z} M_i^{zw}h_{5ij}'
\Sigma_q^x M_j^{qy}   
\nonumber\\
\mbox{}-\frac{tr\Sigma^2}{M_P^2}\frac{1}{8}\epsilon_{uvwxy} H_2^u M_i^{vw}
h_{6ij}M_j^{xy}
+\frac {1}{M_P^2} H_{1x}\Sigma_y^{2x} M_{zi}k_{3ij}M_j^{yz}
\nonumber\\
\mbox{}+\frac {1}{M_P^2} H_{1x}M_{yi} \Sigma_z^{2y} k_{4ij}M_j^{xz}
+ \frac {1}{M_P^2} H_{1x}\Sigma_y^x M_{ui} \Sigma_z^{u} k_{5ij}M_j^{yz}
\nonumber\\
\mbox{}+\frac{tr\Sigma^2}{M_p^2} H_{1x}M_{yi}k_{6ij}M_j^{xy}
\end{eqnarray}

\begin{eqnarray}
W_6= \frac {1}{M_P^3} H_{1x}\Sigma_y^{3x} M_{zi}k_{7ij}M_j^{yz}
+ \frac {1}{M_P^3} H_{1x}M_{yi} \Sigma_z^{3y} k_{8ij}M_j^{xz}
+ \frac {1}{M_P^3} H_{1x}\Sigma_y^{2x} M_{ui} \Sigma_z^{u} k_{9ij}M_j^{yz}
\nonumber\\
\mbox{}+\frac {1}{M_P^3} H_{1x}\Sigma_y^{x} M_{ui} \Sigma_z^{2u} k_{10ij}
M_j^{yz} +\frac{tr\Sigma^3}{M_P^3} H_{1x}M_{yi}k_{11ij}M_j^{xy}
\nonumber\\
+\frac{tr\Sigma^2}{M_P^3}(
 H_{1x}\Sigma_y^x M_{zi}k_{12ij}M_j^{yz}
+  H_{1x}M_{yi} \Sigma_z^y k_{13ij}M_j^{xz})
\end{eqnarray}
After spontaneous breaking of the GUT symmetry when
 $<\Sigma>=M(2,2,2,-3,-3)$, eqs(1-5) create textures.Thus at the 
GUT scale one has

\begin{eqnarray}
W_{eff} 
&=&(-M_{H3}H_{1a}H_2^a+ H_{1a}l_{\alpha 
i}B^E_{ji}q^{\alpha}_{aj}+\epsilon_{abc} H_{1a}d^c_{bi}
B^D_{ji}u^c_{cj}\nonumber \\
& &\mbox{}+H^a_2 u^c_{ai}B^U_{ji}e^c_j +\epsilon_{abc}H_2^au_{bi}
C^U_{ji}d_{cj})\nonumber\\
& &\mbox{}+H_{1\alpha}l^{\alpha}_iA^E_{ji}e^c_j+H_{1\alpha}d^c_iA^D_{ji}q^{\alpha}_j+
H_2^{\alpha}u^c_iA^U_{ji}q_{aj}
\end{eqnarray}
Here $A^E,A^D,A^U$ are the textures in the Higgs doublet sector and 
 $B^E,B^D,B^U$ and $C^U$ are the textures in the Higgs triplet sector.
They contain a hierarchy of mass scales since $W_n$ contributes 
 terms of  $O(M/M_P)^{n-3}$ to the textures. Next we impose on eq(2-5) 
the condition that $A^E, A^D$,and $A^U$, be the GJ textures,
i.e.,

\begin{equation}
A^E=\left(\matrix{0&F&0\cr
                  F&-3E&0\cr
                  0&0&D\cr}\right),
A^D=\left(\matrix{0&Fe^{i\phi}&0\cr
                  Fe^{-i\phi}&E&0\cr
                  0&0&D \cr}\right),
A^U=\left(\matrix{0&C&0\cr
                  C&0&B\cr
                  0&B&A\cr}\right)
\end{equation}

\noindent
The texture zeros of eq(7) are generated provided, 

\begin{eqnarray}
h_{ij}=h \delta_{i3}\delta_{j3},
h_{kij}=h_k (\delta_{i2}\delta_{j3}+\delta_{i3}\delta_{j2}) ; k=1,2 \nonumber\\
h_{\it lij}=h_{\it l}(\delta_{i1}\delta_{j2}+\delta_{i2}\delta_{j1});
\it l=3,4,5,5',6\nonumber \\
k_{ij}=k \delta_{i3}\delta_{j3} ;~k_{qij}=k_q \delta_{i3}\delta_{j3} , q=1,2 ;
k_{pij}=k_p \delta_{i2}\delta_{j2} , p=3,..,6\nonumber\\
k_{rij}=(k_r\delta_{i1}\delta_{j2}+k_r^*\delta_{i2}\delta_{j1}) ; r=7,..,13 
\end{eqnarray}
\noindent

\noindent
$A^E,A^D,A^U$  constructed in the above fashion
contain the desired hierarchies in powers of $\epsilon\equiv M/M_P$.
In the up quark sector $A\sim h$, $B\sim \epsilon h_k$, 
$C\sim \epsilon^2 h_{\it l}$, and we find that A,B,C have the correct 
hierarchical orders   when $\epsilon \sim O(1/50)$ and 
 $h, h_k, h_{\it l}\sim O(1)$. In the down quark lepton sector we have 
 $D\sim k+\epsilon k_q$, $E\sim \epsilon^2 k_p$, and $F\sim \epsilon^3 k_r$, 
and we find that D,E,F have the correct hierarchical orders
with $k=0$, $k_q, k_p,k_r\sim O(1)$.

	There is a weakness,however, in the above approach which we 
now illustrate. It resides in the lack of a full determination of the 
the coupling parameters that appear in the higher dimensional operators of 
eqs(3-5) even with the imposition of the GJ texture constraints.Consider 
the up quark sector first. Here the (33) elements of $B^U$ and $C^U$ are 
uniquely fixed since there is one parameter(h) and one GJ texture 
constraint.For the (23+32) elements there are two parameters $(h_1,h_2)$ and 
one GJ texture constraint. However, fortuitously $(h_1,h_2)$ enter in the 
exact same combination both in $A^U$ and in $B^U, C^U$, and so 
the (23+32) elements of $B^U$ and $C^U$ are again uniquley determined.
However,for the (12+21) elements,one has five coupling constants
($h_3,h_4,h_5,h_5',h_6$) and one GJ texture constraint.Thus 
there is a four parameter arbitrariness here. In the down quark and 
lepton sector, the determination of the (33) element
in $A^E,A^D$ involves the parameters $k_1,k_2$. However, the $k_2$ term 
spoils the $b/\tau$ unification at the GUT scale, so we set $k_2=0$ in 
conformity with the GJ texture constraints of eq(7). With
this constraint the down quark lepton system is uniquely determined in the
(33) element and thus the elements $B_{33}^E$ and $B_{33}^D$ are uniquely 
determined. In the (22) element, there are four coupling constants($k_1,k_2,
k_3,k_4$) and two constraints , one from $A^E$ and the other from $A^D$,
which leave us with a two parameter arbitrariness.  Finally,
in the (12+21) elements, one has seven parameters ( $k_7,..,k_{13}$) and two 
constraints, one each from $ A^E$ and $A^D$. Thus there is a five parameter
arbitrariness in the system at this level. The textures in the Higgs triplet 
sector are given by 

\begin{equation}
B^U=\left(\matrix{0&{4\over 9}C+\Delta_{12}^U&0\cr
                  {4\over 9}C+\Delta_{21}^U&0&-{2\over 3}B\cr
                  0&-{2\over 3}B&A\cr}\right),
C^U=\left(\matrix{0&{4\over 9}C+\Delta_{12}^{U'}&0\cr
                  {4\over 9}C+\Delta_{21}^{U'}&0&-{2\over 3}B\cr
                  0&-{2\over 3}B&A\cr}\right)
\end{equation}
where $\Delta_{12}^U, \Delta_{12}^{U'}$ are given by 

\begin{equation}
\Delta_{12}^U=\epsilon^2(\frac {25}{6} h_4 + \frac {50}{3}h_6-
\frac{50}{9}h_5')
\end{equation}

\begin{equation}
\Delta_{12}^{U'}=\epsilon^2(\frac {25}{6} h_4 + \frac {50}{3}h_6
+\frac {25}{36} h_5')
\end{equation}
and 
\begin{equation}
B^E=\left(\matrix{0&(-{19\over 27}+e^{i\phi})F+\Delta_{12}^E&0\cr
  (-{19\over 27}+e^{-i\phi})F+\Delta_{21}^E&{16\over 3}E+\Delta_{22}^E&0\cr
                  0&0&{2\over 3}D\cr}\right) 
\end{equation}

\begin{equation}
B^D=\left(\matrix{0&-{8\over 27}F+\Delta_{12}^D&0\cr
       {-8\over 27}F+\Delta_{12}^D&-{4\over 3}E+\Delta_{22}^D&0\cr
                  0&0&-{2\over 3}D\cr}\right)
\end{equation}
where $ \Delta_{12}^E,\Delta_{12}^D$ are given by  

\begin{equation}
\Delta_{12}^{E}=\epsilon^3( -25 (k_9 +k_{10}) + \frac {350}{9}
k_{11}-\frac {100}{3}(k_{12}+k_{13}))
\end{equation}

\begin{equation}
\Delta_{12}^{D}= \epsilon^3(-\frac {350}{9}
k_{11}+\frac {100}{3}(k_{12}+k_{13}))
\end{equation}
and where $ \Delta_{22}^E,\Delta_{22}^D$ are {given by  

\begin{equation}
\Delta_{22}^{E}=\epsilon^2(25 k_5 -\frac {50}{3} k_6)
\end{equation}
\begin{equation}
\Delta_{22}^{D}=\epsilon^2(\frac {50}{3}) k_6
\end{equation}

We consider now solutions where all the arbitrary parameters except for
those necessary to satisfy the GJ texture constraints are set to zero.We call 
these the minimal parameter solutions. For $A^U$ 
we find 5 solutions two of which,however, are degenerate in the 
Higgs triplet sector leaving us with only four distinct solutions for
 $B^U$ and $C^U$, listed as  
$A_1,..,A_4$ in table1 ( The case $A_5$ is similar to the case
$A_1$ and is not distinct). In the down quark and lepton sector there
are five minimal parameter solutions that give the same (22) GJ texture 
element in $A^E$ and $A^D$ but lead to distinct $B_{22}^E$ and $B_{22}^D$ 
elements and are listed as cases $B_1,..,B_5$  in 
table2. In the (12+21) down quark lepton sector there are
seventeen minimal two parameter
 solutions that give the same ($A_{12}^E,A_{21}^E$),
and  ($A_{12}^D,A_{21}^D$) elements, and lead to distinct values 
for ($B_{12}^E,B_{21}^E$), and  ($B_{12}^D,B_{21}^D$).
These are exhibited in table 3 and the corresponding elements 
$B_{12}^E$ and $B_{12}^D$ are  listed as cases $C_1,...,C_{17}$ in table4.
We find then that there are four minimal parameter solutions that lead
to distinctly different textures in the $H_1^a$ color interactions 
and $5\times 17$ minimal parameter solutions that lead to distinctly 
different textures in the $H_2^a$ color interactions, while  giving the
exact same GJ textures in the Higgs doublet sector.We label these models
 by $A_iB_jC_k$ where i=1,..,4;~j=1,..,5;~k=1,..,17. It can be easily seen 
from tables 1,2 and 4 that 
	a subset of these minimal parameter models, i.e., $A_iB_mC_n$, 
	where i=1,..,4;~m=1,4;~n=1,4,7,8,9,16,17  satisfies the texture 
	sum rule\cite{hierarchy} $A^E+B^E+B^D=A^D$, while the 
	remaining subset violates the sum rule. The source of these violations 
	can be traced to the couplings $k_{5ij}$ in $W_5$ of eq(4) and the 
	couplings $k_{9ij}$ and $k_{10ij}$ in $W_6$ of eq(5).These are 
	the couplings where the $\Sigma$-field appears at more than one
	location in the interaction structure.  
	  
	The analysis given above shows that the textures in the  
	the $H_2^a$ sector have a 4 parameter arbitrariness
	while the textures in the $H_1^a$ sector have a $3+6$ parameter
	arbitrariness. If one integrates out the heavy color
	higgs fields, one finds as usual dimension five operators 
	which are of the type LLLL and RRRR, where L(R) denote 
	chiralities.The LLLL part involves the textures $B^E$ and $C^U$
	while the RRRR part involves the textures $B^D$ and $B^U$.We find
	that each of these parts involves a 4+3+6 parameter arbitrariness.
	Thus the proton lifetime predictions are rendered highly ambiguous 
	in the general case.
	If we make the choice of picking the minimal number of
	parameters to satisfy the texture constraints in the Higgs doublet
	sector ,then one has $4\times 5\times 17$ different possibilities 
	for the LLLL+RRRR dimension five operators as exhibited in tables
	1,2 and 4. Thus each of the $4\times 5\times 17$ $A_iB_jC_k$ models will
	lead to its own set of proton decay predictions. From tables 1, 2 and
	4 we see that the (12) and (22) texture elements  
	show large variations which will translate into significant 
	variations for proton decay lifetimes.As pointed out in 
	ref\cite{hierarchy} there is also the additional feature  that 
	the CP violating phase enters the Higgs triplet textures. This phase
	influences proton decay lifetimes and decay signatures as it 
	enters prominently in the LLLL and  in the RRRR dimension five operators.
	 
	 Thus the textures derived  from the most general expansions based
	on higher dimensional operators do not lead to a predictive
	theory for proton decay. The arbitrariness encountered arises due 
	to the possibility of writing an operator of higher 
	dimensionality in several different ways due to the several ways one
	can  
	contract the indices.This kind of arbitrariness is  not expected to
	be removed by the so called horizontal symmetries since the nature and 
	number of fields in each configuration is the same for all the terms
	at a given level of dimensionality.  One needs more constraining 
	principles to reduce the arbitrariness in the theory.

	In ref\cite{hierarchy} a model was proposed which reduces the 
	arbitrariness 	
	encountered above by deriving  the higher dimensional operators
	from a dynamical postulate. The proposed model extends  
	supergravity unification to include an exotic sector with couplings
	to both the visible and the 
	hidden sectors.After spontaneous breaking of supersymmetry 
	exotic matter becomes superheavy and its elimination leads 
	to a well defined set of higher dimensional operators.It is then 
	shown that the assumption of an exotic sector belonging to
	 the simplest vector like representation leads to 
	predictive textures in the Higgs triplet sector when the 	
	texture constraints in the Higgs doublet sector are imposed.
	The model of ref \cite{hierarchy} is the case 
	$A_1B_1C_1$ in the notation of tables1,2
	and 4 ( corresponding to  $\Delta_{12}^U,\Delta_{12}^{U'},
	\Delta_{12}^E,\Delta_{12}^D, \Delta_{22}^E$ and $\Delta_{22}^D$ all
	equal to zero) and leads to predictive proton decay lifetime and
	decay signatures.

\section*{Acknowledgements}
This research was supported in part by NSF grant number PHY--19306906 and 
at the Institute for Theoretical Physics in Santa Barbara under grant number 
PHY94-07194.

\newpage
\begin{center} \begin{tabular}{|c|c|c|c|c|c|}
\multicolumn{6}{c}{Table~1:  Evaluation of $\Delta_{12}^U$ and $\Delta_{12}
^{U'}$ and $B_{12}^U$ and $C_{12}^U$ } \\
\multicolumn {2}{r}
{for minimal parameter models. } \\
\hline
Model &$(h_3,h_4,h_5,h_5',h_6)$  &$\Delta_{12}^U$  &$\Delta_{12}
^{U'}$ & $B_{12}^U$ & $C_{12}^U$\\
\hline
$A_1$ &$(\frac{1}{9},0,0,0,0)C$  &$0$
& 0& $\frac {4}{9}C$ & $\frac {4}{9}C$ \\
\hline
$A_2$ &$(0,\frac{4}{21},0,0,0)C$  & $ \frac {50}{63} C$
&$ \frac {50}{63} C$ & $\frac {26}{21}C$ & $\frac {26}{21}C$\\
\hline
$A_3$ &$(0,0,0,-1,0)C$  &$\frac {50}{9}C$
& $-\frac {25}{36}C$ &$6 C$ & $-\frac {1}{4}C$ \\
\hline
$A_4$ &$(0,0,0,0,\frac{1}{30})C$  &$ \frac{5}{9}C$
&$ \frac{5}{9}C$ & $C$ & $C$\\
\hline
$A_5$ &$(0,0,-\frac {4}{9},0,0)C$ &$ 0$
& $0$ & $\frac {4}{9}C$ & $\frac {4}{9}C$\\
\hline
\end{tabular} 
\end{center}

\begin{center} \begin{tabular}{|c|c|c|c|c|c|}
\multicolumn{6}{c}{Table~2:  Evaluation of $\Delta_{22}^E$, $\Delta_{22}
^{D}$, $B_{22}^E$ and $B_{22}^D$  } \\
\multicolumn {2}{r}
{for minimal parameter models. } \\
\hline
Model &$(k_3,k_4,k_5,k_6)$  &$\Delta_{22}^E$  &$\Delta_{22}
^{D}$ & $B_{22}^E$ & $B_{22}^D$\\
\hline
$B_1$ &$(\frac{7}{15},-\frac{4}{5},0,0)E$  &$0$
& 0 &$\frac {16}{3}E$ & $-\frac {4}{3}E$\\
\hline
$B_2$ &$(-\frac{1}{15},0,-\frac{4}{15},0)E$  & $ -\frac {20}{3} E$
&$ 0$  &$-\frac {4}{3}E$ & $-\frac {4}{3}E$\\
\hline
$B_3$ &$(0,-\frac {1}{10},-\frac{7}{30},0)E$  &$-\frac {35}{6}E$
& $0$  &$-\frac {1}{2}E$ & $-\frac {4}{3}E$\\
\hline
$B_4$ &$(0,-\frac {4}{5},0,\frac {7}{50})E$  &$ -\frac{7}{3}E$
&$ \frac{7}{3}E$   &$ 3 E$ & $E$ \\
\hline
$B_5$ &$(0,0,-\frac {4}{15},-\frac {1}{50})E$ &$ -\frac {19}{3}E$
& $-\frac{1}{3} E$  &$-E$ & $-\frac {5}{3}E$\\
\hline
\end{tabular} 
\end{center}

\newpage
\begin{center} \begin{tabular}{|c|c|c|}
\multicolumn{3}{c}{Table~3:  Evaluation of $\Delta_{12}^E$ and $\Delta_{12}
^D$ for minimal parameter models. } \\
\hline
$(k_7,k_8,k_9,k_{10},k_{11},k_{12},k_{13})$  &$\Delta_{12}^E$  &$\Delta_{12}
^D$\\
\hline
$ \left(\one,\two,0,0,0,0,0 \right)F$  &$0$
& 0 \\
\hline
$ \left(\three,0,\four,0,0,0,0 \right)F$  & $ \fourp $
& 0 \\
\hline
$ \left(\term1,0,0,\k10,0,0,0 \right)F$  &$\kx F$
& 0 \\
\hline
$ \left(\five,0,0,0,0,0,\six \right)F$  &$ \sixp$
& $-\sixp$ \\
\hline
$ \left(0,\seven,\eight,0,0,0,0 \right)F$  &$ \nine$
& $0$ \\
\hline
$ \left(0,\ten,0,\eleven,0,0,0 \right)F$  &$ \twelve$
& $0$ \\
\hline
$ \left(0,\thirteen,0,0,\fourteen,0,0 \right)F$  &$ -\fifteen$
& $\fifteen$ \\
\hline
$ \left(0,\sixteen,0,0,0,\seventeen,0 \right)F$  &$ \eighteen$
& $-\eighteen$ \\
\hline
$ \left(0,\nineteen,0,0,0,0,\twenty \right)F$  &$ \twentyone$
& $\twentytwo	$ \\
\hline
$ \left(0,0,\tthree,\tfour,0,0,0 \right)F$  &$ \frac {25}{27}F$
& $0$ \\
\hline
$ \left(0,0,\tfive,0,\tsix,0,0 \right)F$  &$ \tseven $
& $\teight$ \\
\hline
$ \left(0,0,\tnine,0,0,\tten,0 \right)F$  &$ \televen $
& $\ttwelve$ \\
\hline
$ \left(0,0,0,\sone,\stwo,0,0 \right)F$  & $ \sthree $
& $\sfour$ \\
\hline
$ \left(0,0,0,\sfive,0,\ssix,0 \right)F$  & $ \sseven $
& $\ssevenp$ \\
\hline
$ \left(0,0,0,\vone,0,0,\vtwo \right)F$  & $ \vthree $
& $\vfour$ \\
\hline
$ \left(0,0,0,0,\vfive,0,\vsix \right)F$  & $ -\vseven $
& $\vseven$ \\
\hline
$ \left(0,0,0,0,0,\veight,\vnine \right)F$  & $ \vten $
& $-\vten$ \\
\hline
\end{tabular} 
\end{center}

\newpage
\begin{center} \begin{tabular}{|c|c|c|}
\multicolumn{3}{c}{Table~4:  Evaluation of $B_{12}^E$ and $B_{12}
^D$ for minimal parameter models. } \\
\hline
Model  &$B_{12}^E$  &$B_{12}
^D$\\
\hline
$C_1$ &$(-\frac{19}{27}+\eiphi)F$
& $-\frac {8}{27}F$ \\
\hline
$C_2$   & $(-\frac{4}{27}+\frac {4}{9}\eiphi)F$
&  $-\frac {8}{27}F$  \\
\hline
$C_3$ 
&$(\frac{26}{27}-\frac {2}{3}\eiphi)F$
&  $-\frac {8}{27}F$  \\
\hline
$C_4$ & $(-\frac{13}{27}+\frac {7}{9}\eiphi)F$ 
& $(-\frac{14}{27}+\frac {2}{9}\eiphi)F$ \\
\hline
$C_5$  & $-(\frac{13}{9}+\frac {3}{2}\eiphi)F$
& $-\frac{8}{27}F$ \\
\hline
$C_6$ & $(-\frac{1}{3}+\frac {9}{4}\eiphi)F$
& $-\frac{8}{27}F$ \\
\hline
$C_7$  &$ -F$
& $\eiphi F$ \\
\hline
$C_8$ & $(-\frac{13}{21}+\frac {9}{7}\eiphi)F$ & 
$-(\frac{8}{21}+\frac {2}{7}\eiphi)F$  \\
\hline
$C_9$  &$-F$ & $\eiphi F $ \\
\hline
$C_{10}$  & $(\frac{2}{9}+\eiphi)F$
& $-\frac{8}{27}F$ \\
\hline
$C_{11}$ &
$-(\frac{2}{3}+\frac {1}{3}\eiphi)F$
& $(\frac{2}{9}+\frac {7}{9}\eiphi)F$ \\
\hline
$C_{12}$  &$ \frac{2}{3}\eiphi F $
& $-(\frac{4}{9}+\frac {2}{9}\eiphi)F$\\
\hline
$C_{13}$ & $(2-3\eiphi)F$
& $(-\frac{4}{3}+\frac {7}{3}\eiphi)F$ \\
\hline
$C_{14}$ &$\frac{2}{3}F$
& $-\frac {2}{3}\eiphi F$ \\
\hline
$C_{15}$  & $\frac{3}{2}\eiphi F$
& $(-\frac{4}{9}+\frac {1}{3}\eiphi)F$ \\
\hline
$C_{16}$  & $-F$ & $\eiphi F$ \\
\hline
$C_{17}$ & $(-\frac{1}{3}+\eiphi)F$
& $-\frac {2}{3}F$ \\
\hline
\end{tabular} 
\end{center}

\end{document}